# Vapor phase epitaxy of antimonene-like nanocrystals on germanium by an MOCVD process

*Raimondo Cecchini,*[*a] *Christian Martella,*[a] *Claudia Wiemer,*[a] *Alessio Lamperti,*[a] *Alberto Debernardi,*[a] *Lucia Nasi,*[b] *Laura Lazzarini,*[b] *Alessandro Molle,*[a] *and Massimo Longo*[*a]

[a]CNR-IMM, Unit of Agrate Brianza, Via C. Olivetti 2, 20864 Agrate Brianza (MB), Italy;

[b]CNR-IMEM, Parco Area delle Scienze, 37/a, 43124, Parma, Italy.

*Corresponding authors e-mail: massimo.longo@mdm.imm.cnr.it, raimondo.cecchini@mdm.imm.cnr.it; Phone: +39 039603 5085, Fax: +39 039 6881175; CNR-IMM, Unit of Agrate Brianza, via C. Olivetti 2, Agrate Brianza (MB), Italy.

**Abstract.**

Synthetic two-dimensional (2D) mono-elemental crystals, namely X-enes, have recently emerged as a new frontier for atomically thin nanomaterials with on-demand properties. Among X-enes, antimonene, the *β*-phase allotrope of antimony, is formed by atoms arranged in buckled hexagonal rings bearing a comparatively higher environmental stability with respect to other players of this kind. However, the exploitation of monolayer or few-layer antimonene and other 2D materials in novel opto-electronic devices is still hurdled by the lack of scalable processes. Here, we demonstrated the viability of a bottom-up process for the epitaxial growth of antimonene-like nanocrystals (ANCs), based on a Metal-Organic Chemical Vapor Deposition (MOCVD) process, assisted by gold nanoparticles (Au NPs) on commensurate (111)-terminated Ge surfaces. The growth mechanism was investigated by large- and local-area microstructural analysis, revealing that the etching of germanium, catalyzed by the Au NPs, led to the ANCs growth on the exposed Ge (111) planes. As a supportive picture, ab-initio calculations rationalized this epitaxial relationship in terms of compressively strained *β*-phase ANCs. Our process could pave the way to the realization of large-area antimonene layers by a deposition process compatible with the current semiconductor manufacturing technology.

**Keywords:** 2D materials, antimonene, Xenes, MOCVD, germanium, ab-initio calculations

1. **Introduction.**

Since the discovery of graphene [1], several other monoelemental, two-dimensional (2D) materials or Xenes, have been investigated [2],[3]. Antimony represents one of most promising candidates in this framework. In particular, among all possible allotropes, its rhombohedral structure (*β*-phase), besides high environmental stability [4],[5], should exhibit high carrier mobility and low lattice thermal conductivity [6], as well as a sizable band gap, making it a promising material for future electronic, thermoelectric and spintronic devices [7],[8],[9]. Some studies have focused on the crystal and electron band structure of ultra-thin Sb films, suggesting two-dimensional (2D) topological insulator properties for a few bilayers-thick Sb film deposited on a Si(111)-(6 ×6)Au substrate [10]; similar promising results have been found for a single Sb bilayer grown on $Sb_2Te_3$ and $Bi_2Te_3$ substrates [11] [J. Appl. Phys., 119, 015302 (2016), https://doi.org/10.1063/1.4939281]. Reported growths of monolayer antimonene crystals include domains of lateral sizes limited to a few tens of nm [12],[13]. In general, the observed thickness of single-crystal antimonene domains varies from ~1 to few tens of nanometers, and tends to increase with the lateral size of the domains [4]. Nevertheless, even in this thickness regime, *β*-antimonene exhibits promising electrical and optical properties. For example, 5 nm-thick multi-layer antimonene nanoribbons, obtained by a bottom-up plasma-assisted deposition, were found to exhibit a band gap of ~2.03 eV[14]. In Ref.[15], a ~17 nm thick polycrystalline antimonene was grown on MoS$_2$ as a potential candidate for contact metal applications.

Within a technology framework, Sb films with thickness ranging from 3 to 10 nm have recently been reported to show phase change memory functions [16], thus candidating antimonene as an emerging option for single-element memory storage devices.

Despite the deposition process variability, directions for a large-area scalability for the antimonene growth with technology-oriented methods are still undefined [17]. There are only a few examples of antimonene flakes produced by mechanical [5] and liquid-phase exfoliation [18]. Similar to other single-element 2D crystals [8], molecular beam epitaxy (MBE) is the reference technique for producing epitaxial antimonene layers with high structural quality [19],[20],[21][22] [Nano Lett. 2018, 18, 2133−2139, DOI:10.1021/acs.nanolett.8b00429], [23] [2D Mater. 6 (2019) 045028, https://doi.org/10.1088/2053-1583/ab33ba]. Due to the intrinsic limitations of all the above-mentioned approaches, consideration of alternative methods aiming at a scalable antimonene growth is mandatory for its technology exploitation. Metal-Organic Chemical Vapor Deposition (MOCVD) processes leading to epitaxy can be a viable solution for this urgent issue, as they combine the advantages of the bottom-up, selective and conformal growth inherent to all CVD processes, to large-area synthesis, reproducibility and ease-of-transfer to industrial applications [24],[25]; it is worth mentioning here some of the most relevant advances allowed by the use of MOCVD in the field of epitaxial III-V-nitride compounds [26] [*Applied Surface Science 471 (2019) 231–238, https://doi.org/10.1016/j.apsusc.2018.12.011*] [27] *Applied Surface Science 518 (2020) 146218, https://doi.org/10.1016/j.apsusc.2020.146218*, [28] *Nano* Energy 69 (2020) 104427, https://doi.org/10.1016/j.nanoen.2019.104427], and, in particular, in relation with 2D materials, such as MoS$_2$ [29][30] [Nano Research, 2019, 12(10): 2646–2652, https://doi.org/10.1007/s12274-019-2502-9, Nature 520, 656-660(2015), doi:10.1038/nature14417 ] and boron nitride [31] [Scientific Reports, (2019) 9:5736, https://doi.org/10.1038/s41598-019-42236-4].

On the other hand, the only previous result about elemental growth of Sb by MOCVD reported the formation of 3D crystals in relatively rough films, owing to coalescence [32].

As for other 2D materials, the choice of an appropriate substrate represents a key factor to attain large-area epitaxial growth and can be inspired by lattice match arguments [20]. In this respect, based on previous MBE results [33], we selected single-crystalline germanium as the substrate bearing the antimonene epitaxy and displaying compatibility with a semiconductor device process flow; at the same time, we propose a new MOCVD-based method for the epitaxial growth of thin (5 – 50) nm, antimonene-like nanocrystals (ANCs) with potential for wafer-scale processing. The related growth mechanism, assisted by gold nanoparticles (NPs) dispersed on the Ge surface, and the structural properties of the grown ANCs were investigated by a wealth of structure- and chemistry-sensitive techniques, including Scanning Electron Microscopy (SEM), Raman spectroscopy, X-Ray Diffraction (XRD) and Transmission Electron Microscopy (TEM). Ab-initio simulations were also compared with experimental data of ANCs on Ge to elucidate the microscopic mechanisms responsible for the epitaxy to occur.

## 2. Experimental.

The growth of ANCs was performed on both p-doped and intrinsic Ge (111) and (001) substrates. The substrates were first treated by immersion in an HF 5% solution for 30 s and rinsed in deionized water to remove the native oxide layer. For the trials with Au NPs, a solution of HF 1% and Au NPs with 10 nm of diameter (British Bio Cell Company ®) was drop casted on the substrates and left for 30 s to have deposition of the NPs on the Ge surface. The substrates were then rinsed in deionized water and blown dry (more details on this procedure in Ref. [34]). The substrates were then loaded into an Aixtron® AIX 200/4 reactor for the MOCVD growth. The Sb precursor was antimony trichloride ($SbCl_3$), provided by Air Liquide®, which was contained in bubblers and transported to the MOCVD reactor by ultra-purified $N_2$ gas [35]. Different MOCVD process conditions were tested. The $SbCl_3$ partial pressure and the reactor total pressure were varied in the range of $(0.9 - 5.6) \times 10^{-3}$ mbar and (100 − 300) mbar, respectively. The MOCVD reactor temperature was set at 325 °C. Growth time was in the range of 30 to 120 min. For the three ANCs samples discussed in the text and labelled N1, N2 and N3 (Fig. 1), the growth was performed on Ge (001) at the same reactor temperature (325 °C) and pressure (300 mbar) but different precursor partial pressures/growth times: N1, SbCl$_3$ partial pressure = $2.8\times10^{-3}$ mbar, growth time = 120 min; N2, SbCl$_3$ partial pressure = $9.0\times10^{-4}$ mbar, growth time = 120 min; N3, SbCl$_3$ partial pressure = $1.7\times10^{-3}$ mbar, growth time = 60 min.

The morphology, microstructure and composition of the ANCs were analyzed by a Zeiss® Supra 40 Scanning Electron Microscopy (SEM). Transmission Electron Microscopy (TEM) was performed using a JEOL 2200FS microscope working at 200 kV equipped with an Energy Dispersive X-ray Spectrometer (EDX), a high-angle annular dark-field (HAADF) detector and an in-column energy (Omega) filter. EDX maps were obtained in scanning TEM (STEM) mode. Raman characterization of the ANCs was performed in a backscattering configuration employing a Renishaw Invia spectrometer, equipped with the 633 nm (1.96 eV) line of a He-Ne laser. The laser radiation was focused on the sample by means of a 50x Leica objective (0.75 numerical aperture), maintaining the incident laser power below 1 mW to avoid sample damage. X-ray diffraction experiments were conducted with a laboratory instrument (HRXRD IS2000) equipped with a sealed Cu K$_\alpha$ tube, a four-circle goniometer a single point (NaI) detector and a position sensitive detector able to collect data on 120°. Precise alignment on the sample surface was performed with the scintillator detector, and the acquisition of the diffracted intensity was performed with the position sensitive detector.

First principles simulation were performed by plane-waves pseudopotential techniques, within the framework of density functional perturbation theory, as implemented in quantum espresso package [36],[37]. The ground state was computed with the PBE [38] generalized gradient approximation for exchange-correlation functional. The norm-conserving pseudopotential and periodic boundary conditions were used. Electronic wave-functions were expanded on a plane-wave basis set with a kinetic energy cutoff of 80 Ry.

3. Results and Discussion

Typical results of ANCs growth on Ge substrates with [111] and [001] crystallographic orientations, pretreated with an HF solution and with the dispersion of Au NPs, are illustrated in Fig. 1.

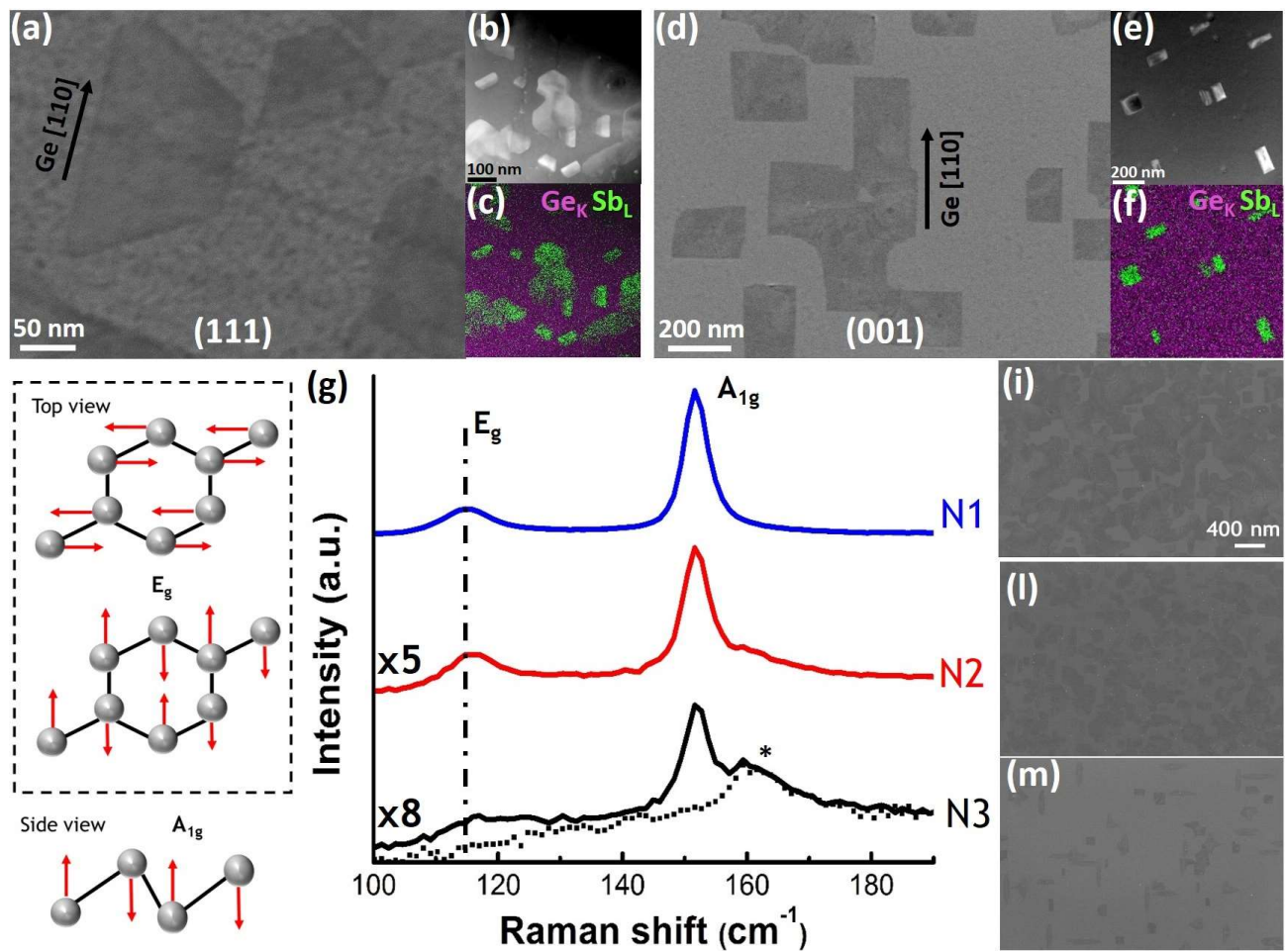

**Fig. 1** SEM plan-view of ANCs grown on HF-treated (a) Ge (111) (corresponding (b) STEM-HAADF image and (c) EDX map) and (d) Ge (001) substrates (corresponding (e) STEM-HAADF image and (f) EDX map), with Au NPs and the same MOCVD process conditions (reactor temperature = 325 °C, reactor pressure = 100 mbar, $SbCl_3$ partial pressure = $1.7 \times 10^{-3}$ mbar, growth time = 120 min). (g) Evolution of the Raman spectra of three ANCs samples: N1, N2 and N3, with different thicknesses (see Experimental method for details), for which the plane-view SEM images are shown in (i), (l) and (m), respectively. The peak at ~160 cm$^{-1}$ marked with * in sample N3 is from the substrate (note that the dotted spectrum refers to a bare germanium substrate). The top and side view sketches of the antimonene $E_g$ and $A_{1g}$ Raman modes are also reported on the left of the Raman spectra.

ANCs appeared to self-organize in domains with sizes of a few hundreds of nm on both Ge (111) and Ge (001) surfaces (darker areas in Fig. 1(a) and (d)). The formation of Sb domains was confirmed by Energy-Dispersive X-ray Spectroscopy (EDX) maps obtained on back-thinned samples (Fig. 2(c) and (f)) and cross-sectional specimens (Fig. S1.1 of supplementary material). The TEM cross-sections analysis allowed to reveal that: i) on both Ge (111) and Ge (001) the growth is accompanied by the etching of the Ge substrate (Fig. S1.1 of supplementary material); ii) the Sb domains grow inside the etched regions. This explains why square/rectangular ANCs domains form on Ge (001), while polygonal ANCs domains form on Ge (111) substrates [39].

We investigated the samples by means of Raman spectroscopy. In particular, we investigated the spectra of samples N1, N2 and N3 (Fig. 1(g)). As evaluated based on SEM and TEM cross-section observations and growth parameters (growth time and precursor partial pressure), the thickness of sample N1 was ~ 40 nm. An analogous estimation for N2 and N3 corresponds to thicknesses ≤ 15 nm for both samples. Assigning a precise value of thickness of the ANCs by SEM and TEM analyses is difficult, especially for the thinnest crystals, because different ANCs on the same substrates can have different thicknesses. The Raman spectrum of the bulk $\beta$-phase Sb is qualified by $E_g$ and $A_{1g}$ Raman active modes positioned at 110.7 cm$^{-1}$ and 149.2 cm$^{-1}$, respectively [4]. The transition from bulk to an antimonene-like regime takes place with a sensible variation in the relevant Raman features, consisting in a blue shift of the $E_g$ and $A_{1g}$ peak frequencies [4]. According to this point of view, all spectra in Fig. 1(g) show the characteristic $E_g$ and $A_{1g}$ peaks of the Raman active modes expected for the $\beta$-phase allotrope (inset of Fig. 1(g)). Moreover, the trend in the peak positions is consistent with the blue-shift expected for the growth of antimonene-like nanocrystals with thickness in the 10 − 20 nm range ($E_g$ and $A_{1g}$ peak positions are: ~117 cm$^{-1}$ ~152 cm$^{-1}$ for N3 sample, ~115.8 cm$^{-1}$ ~ 151.8 cm$^{-1}$ for N2, and ~ 114.7 cm$^{-1}$ ~151.6 cm$^{-1}$ for N1, respectively) [4]. Similar Raman spectra were collected in our samples for ANCs grown on Ge (111) and (001). Without the use of Au NPs, the formation of Sb domains is strongly reduced, particularly on Ge (001).

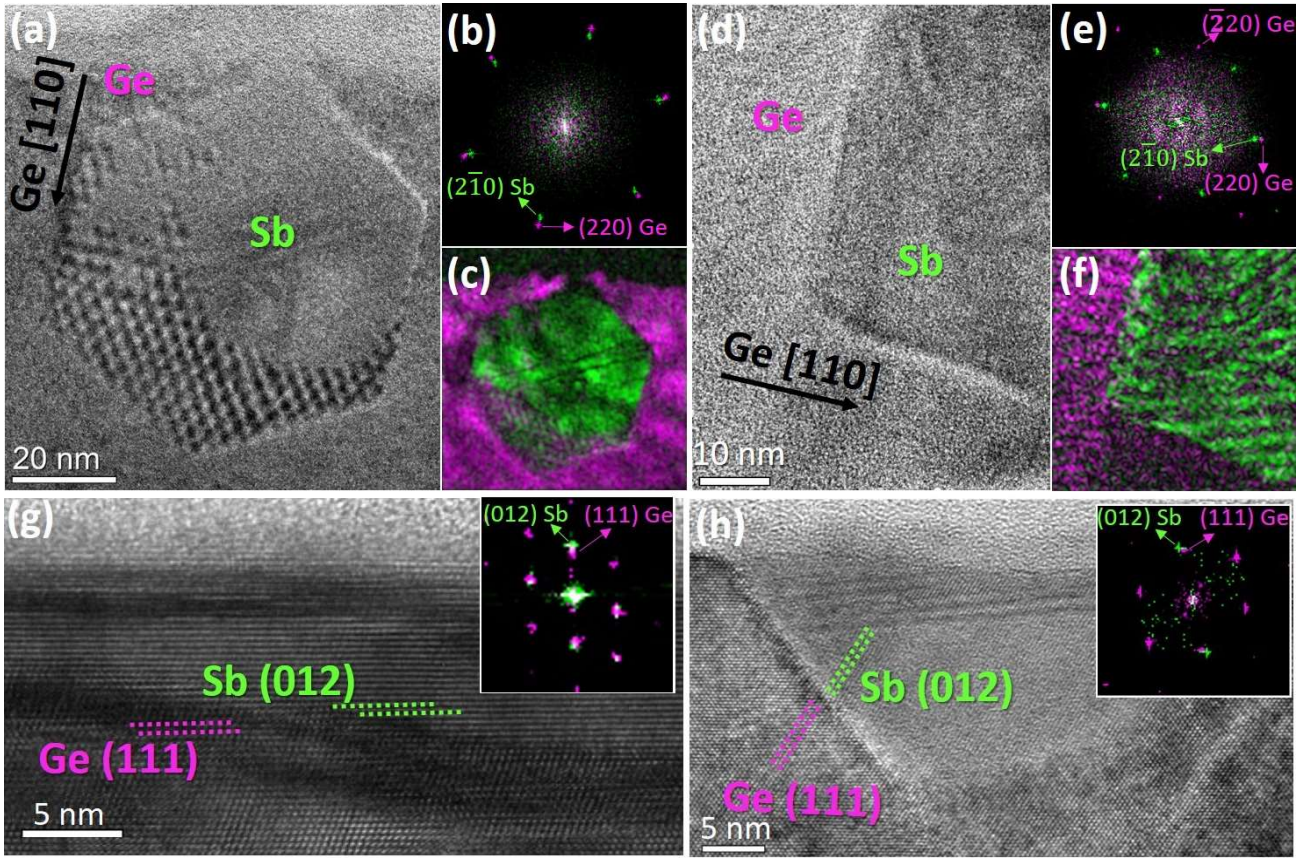

**Fig. 2** (a) Plan-view HR-TEM image of an ANC grown on Ge (111) with Au NPs; (b) FFT of the image in (a); (c) colored inverse FFT map showing the Ge (purple) and Sb (green) phases; (d) plan-view HR-TEM image of an ANC grown on Ge (001) with Au NPs, with the corresponding FFT pattern (e) and inverse FFT color map (f). Cross-section HR-TEM images of an ANC formed on Ge (111) and Ge (001) are displayed in (g) and in (h), respectively, with their corresponding FFT patterns in the insets.

The structural quality of the obtained ANCs, the epitaxial relationship with the substrate and the growth mechanism were then studied by means of High Resolution (HR)-TEM investigations, performed on both plan-view and cross-section specimens (Fig. 2), and on the large scale by XRD measurements (Fig. 3). HR-TEM analysis revealed that the ANCs domains are monocrystalline and epitaxially grown on the substrates. The HR-TEM plan-view images of two antimonene domains on Ge (111) and Ge (001) are shown in Fig. 2(a) and (d), respectively, and the corresponding Fast Fourier transforms (FFTs) in Fig. 2(b) and (e). In the first case, two sets of aligned spots arranged in hexagonal patterns can be observed in the FFT (Fig. 2(b)), each characteristic of the [111] Ge and [001] Sb zone axes, similarly to what reported for the antimonene growth on Ge (111) by MBE [33]. In the second case, the FFT pattern shows the hexagonal spot pattern of the [001] Sb zone axis (green) aligned with the square pattern of [001] Ge (purple) (Fig. 2(e)). Such epitaxial relationship was found in all the observed ANCs domains grown on (001) substrates. The common feature for the observed epitaxial orientations of ANCs with respect to germanium is the alignment of Sb ($2\bar{1}0$) planes with Ge (220) planes (Fig. S1.2 of supplementary material). This might be explained in terms of the relatively low mismatch of the interplanar distances (2.15 Å and 2.0 Å for Sb ($2\bar{1}0$) and Ge (220), respectively).

The cross-section of one ANC domain grown on a cavity of Ge (111) substrate is shown in Fig 2(g): the arrangement of atomic planes observed in the cross section does not always correspond to the epitaxy observed in the plan view ((Fig. 2(a)). This suggests that the irregular and complex shape of the cavities formed on (111) substrates allows different Sb orientations. Conversely, for the cross-section of an ANC grown on (001) Ge substrate (Fig. 2(h)), the corresponding FFT pattern shows that the Sb (012) spots are nearly aligned with the (111) spots of Ge, with a tilt of about 2 degrees. Considering that the interplanar angle between Ge (111) and Ge (001) is 54.7° and the interplanar angle between Sb (012) and Sb (001) is 56.5° (see also XRD analyses in the following), this arrangement is consistent with a (001) Sb out-of-plane orientation, in agreement with the plan view observations (Fig. 2(d)). However, in all cases, the HR-TEM cross section images reveal a structural continuity of Sb (012) planes with Ge (111) planes, as shown in Fig. 2(g) for an ANC domain with a (012) Sb out-of-plane orientation.

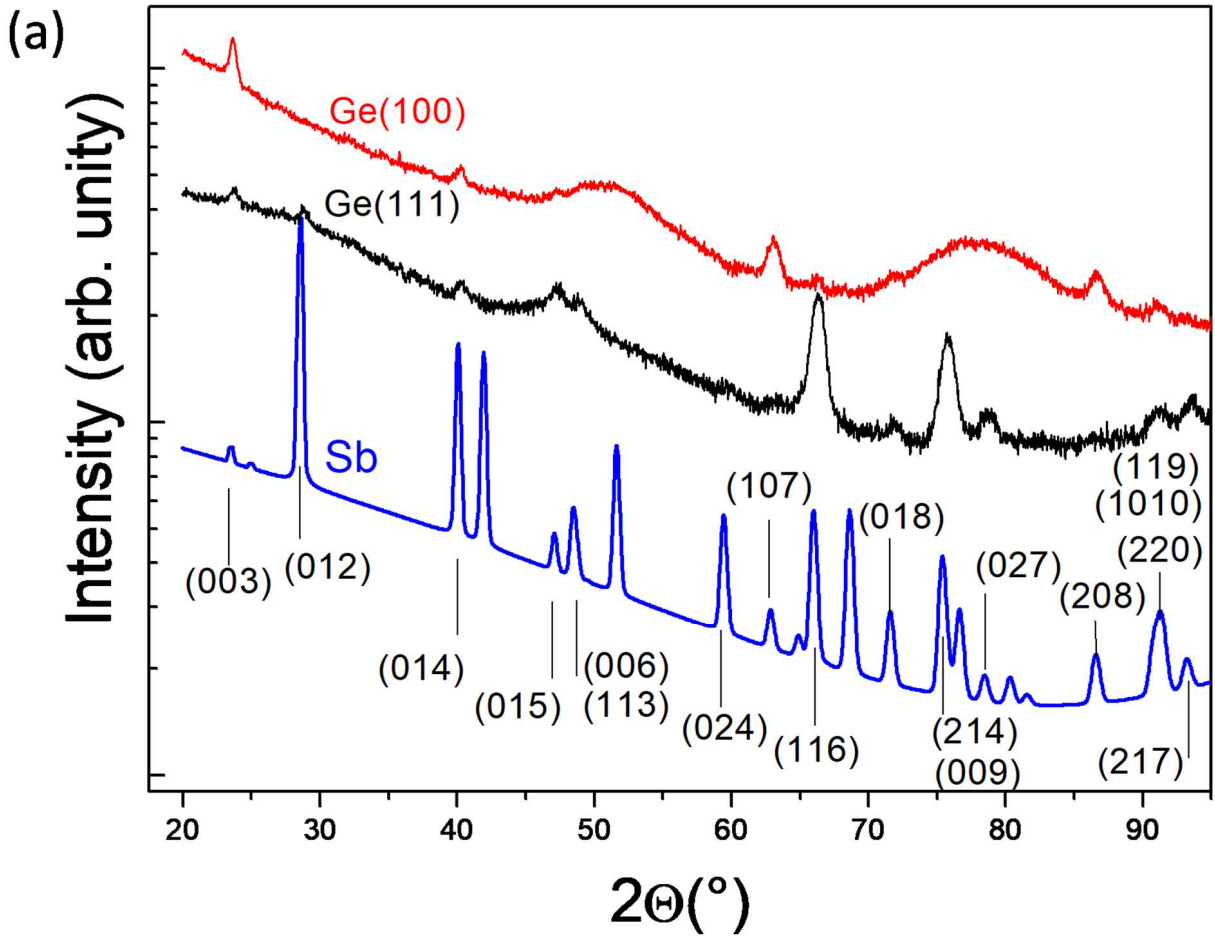

**Fig. 3** Grazing incidence XRD pattterns obtained on ANCs grown on Ge (001), upper (red) curve, and on Ge (111), lower (black) curve. The pattern for powder β-phase Sb [40], is also added for comparison. The peaks present in the XRD patterns are indexed.

Paralleled with TEM, XRD investigations allowed us to study the epitaxial relationship on a macroscopic scale. Studies were performed using both grazing incidence and Bragg Brentano geometries, and by recording complete ϕ scans around asymmetric reflections. Fig. 3 reports the grazing incidence XRD spectra, as obtained for antimonene domains on Ge (001) and Ge (111). The incident radiation enters the samples 20° offset to Ge [011] on Ge (001) and parallel to Ge [011] on Ge (111). All the observed diffracted maxima can be ascribed to the rhombohedric structure of Sb [40]. As compared to the powder diffraction pattern of β-phase Sb (see blued pattern in Fig. 3), the diffractograms present extinguished and enhanced diffracted maxima, which unveil the preferential growth of the domains. In the diffractogram obtained on Ge (111), together with other peaks, all those detected for ANCs grown on Ge (001) are present, although with lower intensity. This analysis shows that one and two preferential orientations of the ANCs were observed on Ge (001) and Ge (111), respectively. Complementary Bragg Brentano analyses (Fig.S2.1 of supplementary material) confirmed the growth of Sb (001) out-of-plane oriented domains on Ge (001), whereas, on Ge (111), only the out-of-plane oriented Sb (012) grains are revealed, the Sb (001) ones (observed in the TEM analysis) being therefore not enough to reach the necessary scattering efficiency to be detected in this geometry. However, complete ϕ scans around asymmetric reflections (Fig.S2.2 of supplementary material), by which a large surface area can be probed and the diffraction efficiency is increased, confirmed that, on Ge (111), both Sb (001) and Sb (012) out-of-plane oriented planes are present. Overall, an epitaxial relationship sets in between the Sb (012) and Sb (001) planes on the one hand, and the Ge (111) and Ge (001) ones on the other hand, with interplanar angles being 56.5° in Sb and 54.7°, respectively. The appearance of Sb (001) out-of-plane oriented planes also on Ge (111) can be ascribed to the structural similarity between the Sb (001) and Sb (012) planes, with similar interatomic distances and arrangements.

The structural relations imposed on the ANCs by the observed epitaxial growth on germanium substrates were also investigated by first principle simulations. In particular, we studied the observed case of $\beta$-phase Sb with the Sb ($2\bar{1}0$) planes aligned with Ge (220) planes, presenting the Ge (001) plane parallel to the Ge (111) (Fig. 2(a); see also Fig. S1.2(b) of supplementary material), thus considering the possibility of epitaxial growth of honeycomb Sb layers on Ge (111). Since the in-plane lattice parameter of the bulk buckled honeycomb structure of $\beta$-phase Sb is ~7% larger than the corresponding in-plane lattice parameter of Ge, an epitaxial Sb layer on the Ge (111) substrate should present a contracted lattice parameter with respect to the bulk one. Epitaxial Sb layers are therefore expected to reduce the strain energy by relaxing the in-plane lattice parameter through the formation of misfit dislocations (see HR-TEM data in Fig. S1.3(c) of supplementary material). If the Sb layers are not fully-relaxed, some residual compressive strain may still be present in them. In our calculations, we set the in-plane lattice parameter of Sb at a desired strained value, while the lattice parameter along the growth direction was free to relax. We computed a blue shift of the Raman modes (see section S3 of supplementary material) that, for a 1% in-plane compression of the lattice-parameter, amounts to about 4 cm$^{-1}$ for both modes. This effect may account, at least in part, for the blue-shift observed in the Raman spectra of the ANCs samples when the film thickness is reduced (we have assumed that, as the sample thickness increases, the in-plane lattice parameter, and thus the Raman frequencies, approach the fully relaxed bulk values). The range of the strain values indicated by first principles simulations is consistent with our XRD and HR-TEM observations (section of S3 of supplementary material).

The HR-TEM, XRD and Raman analyses, together with the differences observed with and without the HF pretreatment and the use of Au NPs, hint at a possible picture of the ANCs growth based on the following facts: 1) local substrate etching is observed concomitantly with ANCs formation; 2) both the HF pretreatment and Au NPs facilitate the growth of the ANCs. Etching of the Ge substrate is likely induced by the Cl$^-$ ions [41],[42],[43] resulting from the molecular dissociation of the SbCl$_3$ precursor. The so-exposed Ge (111) planes could then favor the epitaxial growth of the ANCs [33]. Moreover, the Ge etching process is catalytically boosted by the Au NPs, as one can infer from the well-documented metal-assisted etching of Ge in liquid solutions [44],[45]. In the present case of a gas mixture containing SbCl$_3$, the Au NPs might help the SbCl$_3$ precursor decomposition into ions of Cl and Sb, therefore acting as centers for both Ge etching and Sb growth. A scheme of the described mechanism is shown in Fig. 4.

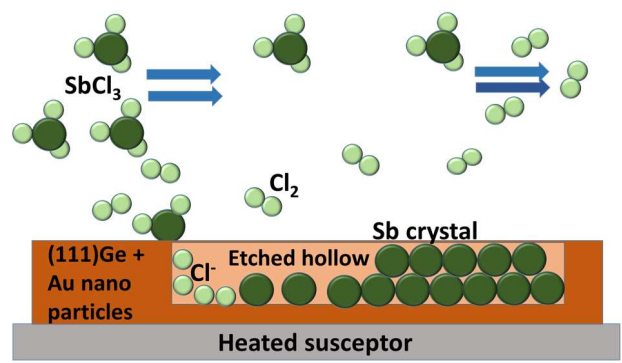

**Fig. 4** Schematic representation of the mechanism of growth of the ANCs: SbCl$_3$ molecules undergo decomposition, assisted by Au NPs, into Sb$^+$ and Cl$^-$ ions. The Cl$^-$ ions etch the Ge substrate; Sb crystals grow on the Ge surfaces exposed by the etching.

In principle, on Ge (111) oriented substrates, the planes suitable for the epitaxial growth of the ANCs are already exposed and there should be no need to generate them by surface etching. However, the growth on the pristine Ge surface could be prevented by the presence of germanium oxide on the surface. While the preliminary HF cleaning should remove oxides from the Ge surface, some re-oxidation (occurring in the elapsed time between the HF etch and the start of the MOCVD process) could inhibit the Sb growth out of the regions where fresh Ge (111) planes are exposed to the Au NP-assisted $Cl^-$ etching during the MOCVD process. Finally, in the SEM plan view images of Fig. 1(e), remains of the Au NPs could be observed in some of the ANCs. On the contrary, we found no evidence of NPs underneath the ANCs in the TEM images. We can suppose that, in some cases, the Au NPs, after favouring the Ge etching, are dissolved by reacting with the Cl ions, to form gaseous gold chloride, according to the reaction described by A. F. Holleman et al. [46] [(A. F. Holleman; Egon Wiberg, Inorganic Chemistry (101 ed.). Academic Press. pp. 1286–1287. ISBN 978-0-12-352651-9].

From the results presented above, it can be expected that control over the germanium substrate etching could allow the growth of large area ANCs. This could be obtained by either a controlled positioning of Au NPs/nanostructures or by replacing the use of Au NPs by, e.g., a plasma-based etching step of the substrates in the MOCVD reactor before the Sb growth.

## 4. Conclusions

In summary, we demonstrated the epitaxial growth of thin (5 – 50 nm) antimonene-like nanocrystals (ANCs) by means of an MOCVD process on single-crystal germanium substrates. Microstructural investigations were used to characterize the obtained Ge/Sb heterostructures and the growth mechanism. The formation of ANCs involves the etch of the germanium substrate (that is induced by $Cl^-$ ions from the precursor molecule dissociation, that is in turn catalysed by Au nanoparticles intentionally dispersed on the substrate prior to growth). This mechanism leads to the exposure of Ge (111) planes, on which antimonene preferentially grows, owing to a compliant epitaxial relationship (based on XRD and HRTEM analyses) and to the accumulation of a compressive strain (according to first principle simulation). Consistently, flat ANCs extended domains were formed when Ge (111) oriented substrates were used. The above illustrated results open up the possibility of further growth process development towards larger areas and lower thickness ANCs with easy industrial transferability, aiming at antimonene exploitation in devices.

**Declaration of Competing Interest**

The authors declare that they have no known competing financial interests or personal relationships that could have appeared to influence the work reported in this paper.

**Acknowledgments**

This project has received funding from the European Union's Seventh Framework Programme (FP7/2007-2013) under grant agreement No 310339 ("SYNAPSE": SYnthesis and functionality of chalcogenide NAnostructures for PhaSE change memories). The authors also thank R. S. R. Gajjela for his help with MOCVD growths and SEM analysis.

**Appendix A. Supplementary material**

The following are the Supplementary data to this article: ****